\documentclass[aps,letterpaper,reprint,showpacs,amsmath,amssymb,superscriptaddress,floatfix]{revtex4-1}

\usepackage{amssymb}
\usepackage{latexsym}
\usepackage[dvips]{graphicx}
\usepackage{epsfig}

\usepackage{dcolumn}
\usepackage{amsmath}
\usepackage{amsfonts}
\usepackage{bm}
\usepackage{color}

\usepackage{hyperref}
\hypersetup{colorlinks,linkcolor=blue,urlcolor=blue,citecolor=blue}

\newcommand{\be}{\begin{equation}}
\newcommand{\ee}{\end{equation}}
\newcommand{\bea}{\begin{eqnarray}}
\newcommand{\eea}{\end{eqnarray}}

\newcommand{\s}{\sigma}

\newcommand{\la}{\langle}
\newcommand{\ra}{\rangle}
\newcommand{\rd}{\mbox{d}}
\newcommand{\ri}{\mbox{i}}
\newcommand{\re}{\mbox{e}}

\newcommand{\Na}{{NaTiSi$_2$O$_6$}}
\newcommand{\Ru}{{NaRuSi$_2$O$_6$}}
\newcommand{\M}{{Na$M$Si$_2$O$_6$}}
\renewcommand{\vec}[1]{{\bm #1}}

\newcommand{\red}[1]{\textcolor{black}{#1}}

\newcommand{\ignore}[1]{}
\begin{document}
\title{Quantum liquid with strong orbital fluctuations: the case of the pyroxene family}
\author{A. E. Feiguin}
\affiliation{Department of Physics, Northeastern University, Boston, MA 02115, USA}
\author{A. M. Tsvelik}
\affiliation{Condensed Matter Physics and Materials Science Division, Brookhaven National Laboratory, Upton, NY 11973, USA}
\author{Weiguo Yin}
\affiliation{Condensed Matter Physics and Materials Science Division, Brookhaven National Laboratory, Upton, NY 11973, USA}
\author{E. S. Bozin}
\affiliation{Condensed Matter Physics and Materials Science Division, Brookhaven National Laboratory, Upton, NY 11973, USA}

\date{\today}

\begin{abstract}
We discuss quasi one-dimensional magnetic Mott insulators from the pyroxene family where spin and orbital degrees of freedom remain tightly bound. We analyze their excitation spectrum and outline the conditions under which the orbital degrees of freedom become liberated so that the corresponding excitations become dispersive and  the spectral weight shifts to energies much smaller than the exchange integral.
  \end{abstract}


\maketitle
\emph{Introduction.---}During the last 30 years a great theoretical effort has been directed at the research on quantum liquids where spin ordering either does not occur or transition temperature is strongly reduced due to fluctuations.   Disordered quantum liquids play an important role in all kinds of theoretic scenarios for exotic matter states. It is well known that quantum fluctuations increase when the symmetry manifold is extended from the ubiquitous SU(2) to a higher symmetry, for instance, SU(N). In practice such extension can occur only when orbital degrees of freedom are included which is difficult since the orbital degeneracy is usually lifted by  the lattice. In this paper we suggest that  magnetic insulators from the so-called pyroxene family may provide a possible path to overcome these difficulties.

Pyroxenes are quasi one-dimensional Mott insulators where spin and orbital degrees of freedom remain tightly bound even at low energies. They compose a very rich class of minerals  with chemical formula $AM$(Si,Ge)$_2$O$_6$ where $A$ is mostly an alkali metal element and $M$ a trivalent metal element. For example, greenish NaAlSi$_2$O$_6$ is a famous Chinese jade called Fei Tsui. The systems with partially filled $d$ shells of the $M$ ions commonly possess nontrivial magnetic properties ranging from antiferromagnetic (AF), ferromagnetic (FM), and spin glassy and likely to be multiferroics, as seen in NaFeSi$_2$O$_6$, LiFeSi$_2$O$_6$, and LiCrSi$_2$O$_6$ \cite{khomskii}. Their crystal structures contain  characteristic \emph{zigzag} chains of edge-sharing $M$O$_6$ octahedra (Fig. {\ref{fig:orbital}). The chains are bridged by the O-Si-O or O-Ge-O bonds, or, in other words, are separated by SiO$_4$ or GeO$_4$ tetrahedra, thus confining the motion of valence electrons to the chains.

 In this paper we discuss pyroxene compounds with $M$ = Ti and Ru, where the lowest $t_{2g}$-orbitals well separated from the $e_{2g}$ ones are occupied either by a single electron (Ti) or a single hole (Ru). At present  only {\Na} has been experimentally studied. It is a relatively simple member of the pyroxene family. Like the V$^{4+}$ ions in the straight-chain system VO$_2$, the Ti$^{3+}$ ions in {\Na} have the $3d^1$ valence electron configuration and undergo the Ti-Ti dimerization upon cooling. In addition, the zigzag chain pattern makes it more apparent that all spin, orbital, and lattice degrees of freedom are active, leading to the so-called two-orbitally assisted Peierls transition \cite{brink,brinkEPL,motome} that generates spin-singlet dimers on the short Ti-Ti bonds \cite{JPSJ} with the spin gap of $\sim 53$ meV \cite{spingap}, rather than a gapless long-range antiferromagnetic (AF) state in VO$_2$. Note that the ordinary spin-Peierls transition seems not to work here because the doubled periodicity is not consistent with the quarter filling of the electronic bands. An early density-functional theory (DFT) study focused on the high-temperature non-dimerized structure (HTS) of {\Na} attributed the spin gap to the spin-one ($S=1$) Haldane type due to the ferromagnetic (FM) Ti-Ti interaction \cite{popovic}. A subsequent DFT calculation with a $U$ correction focused on the low-temperature dimerized structure (LTS) showed that the dominant magnetic interaction was the AF one along the Ti-Ti short bonds, supporting the picture of $S=0$ spin dimers \cite{khomskiiPRL}. However, an outstanding puzzle is that the heat capacity data show the gap $\sim$ 10 meV \cite{JPSJ} suggesting the existence of softer excitations and stronger quantum fluctuations. 

\begin{figure}[t]
   \includegraphics[width=0.45 \textwidth]{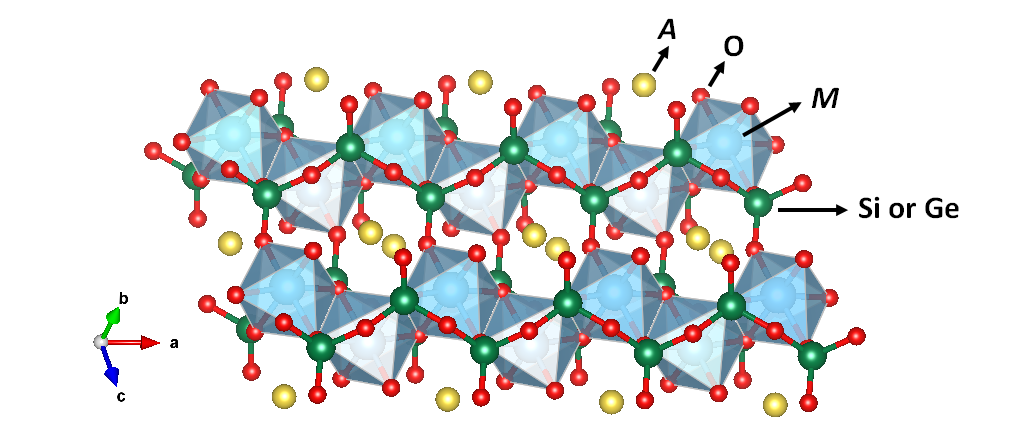}
   \vspace{-0.2cm}
  \caption{\label{fig:orbital}Crystal structure of Na{\it M}Si$_2$O$_6$.}

\vspace{-0.5cm}

\end{figure}


 We approach the problem  using a  combination DFT, analytic, and time-dependent density-matrix renormalization group (DMRG) methods to study their orbital and spin dynamics. We show that stronger quantum fluctuations originate from the involvement of the third $t_{2g}$ orbital, which becomes active when the oxygen-atom-mediated electron hopping integral is comparable to the direct hopping integral between neighboring $M$ atoms \cite{khomskii,mila,jackeli}. 

\emph{Hubbard model and the Sutherland Hamiltonian.---}We start with a microscopic derivation of the three-orbital model Hamiltonian \cite{SI} assuming a single electron or hole occupation of the $t_{2g}$ orbital in this family of Mott insulators. The strong on-site Coulomb interaction $U(N_k-1)^2$ opens a big charge gap $\sim U$ thus preventing direct transitions to states with different occupation number. To get an effective description of the low energy dynamics we have to integrate out the high-energy degrees of freedom as it is done, for instance, in the conventional SU(2) invariant Hubbard model \cite{yin_prb09}. Here, each $M$ cation is coordinated with six O$^{2-}$ anions and the $M$O$_6$ octahedra are edge sharing to form the zigzag chain in the crystallographical $a$ axis (Fig. 1). The five $d$-shell orbitals of the $M$ ion are well separated by the ligand field into the high-energy $e_g$ ($3z^2-r^2$ and $x^2-y^2$) and low-lying $t_{2g}$ ($xy$, $yz$, $zx$) orbitals. The three $t_{2g}$ orbitals are relevant here to the low-energy physics of interest. If one neglects  all factors leading to violation of the SU(6) symmetry, such as the splitting of the $t_{2g}$ orbitals and the Hund's interaction and adopts a diagonal tunneling matrix with identical matrix elements $t$ for all orbitals, the result for $U \gg t$ is the SU(6)-symmetric Sutherland Hamiltonian:
 \bea
H = J\sum_{k}  P^{o,s}_{k,k+1}, ~~ J = \frac{2t^2}{U}\label{Hub}
\eea
where $ P^{o,s} = P^o\otimes P^s$ is the permutation operator acting in 6$\times$6-dimensional space of spin and orbital quantum numbers and $ P^s_{k,k+1}=2\vec{S}_k\cdot\vec{S}_{k+1}+1/2$ and $P^o_{k,k+1} = 2{\vec T}_k{\vec T}_{k+1} + 1/2$, where $S^a, T^a$ are spin and isospin S=1/2 operators acting  on the spin and orbital subspaces, respectively. Model (\ref{Hub}) is integrable, the spectrum consisting of collective orbital and spin excitations is gapless \cite{Sutherland}. The excitations (spinons) are fractionalized, that is they carry quantum numbers of electrons (except the charge one which is gapped), that is  spin 1/2 and orbital indices. Results for the spin spectral function are presented on Fig. \ref{SU6}a.

\begin{figure}[t]
\includegraphics[width=0.45 \textwidth]{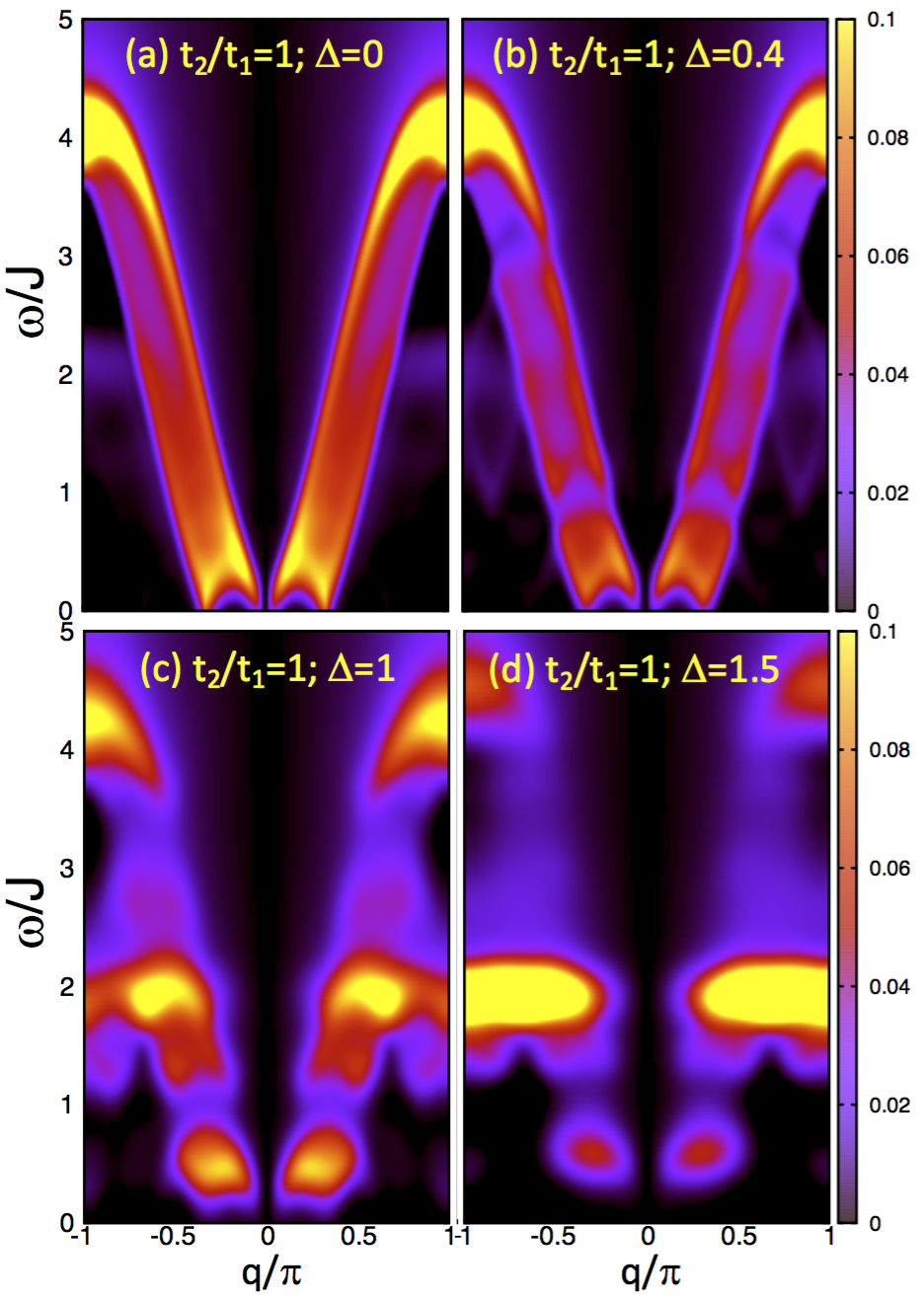}
\vspace{-0.2cm}
 \includegraphics[width=0.45 \textwidth]{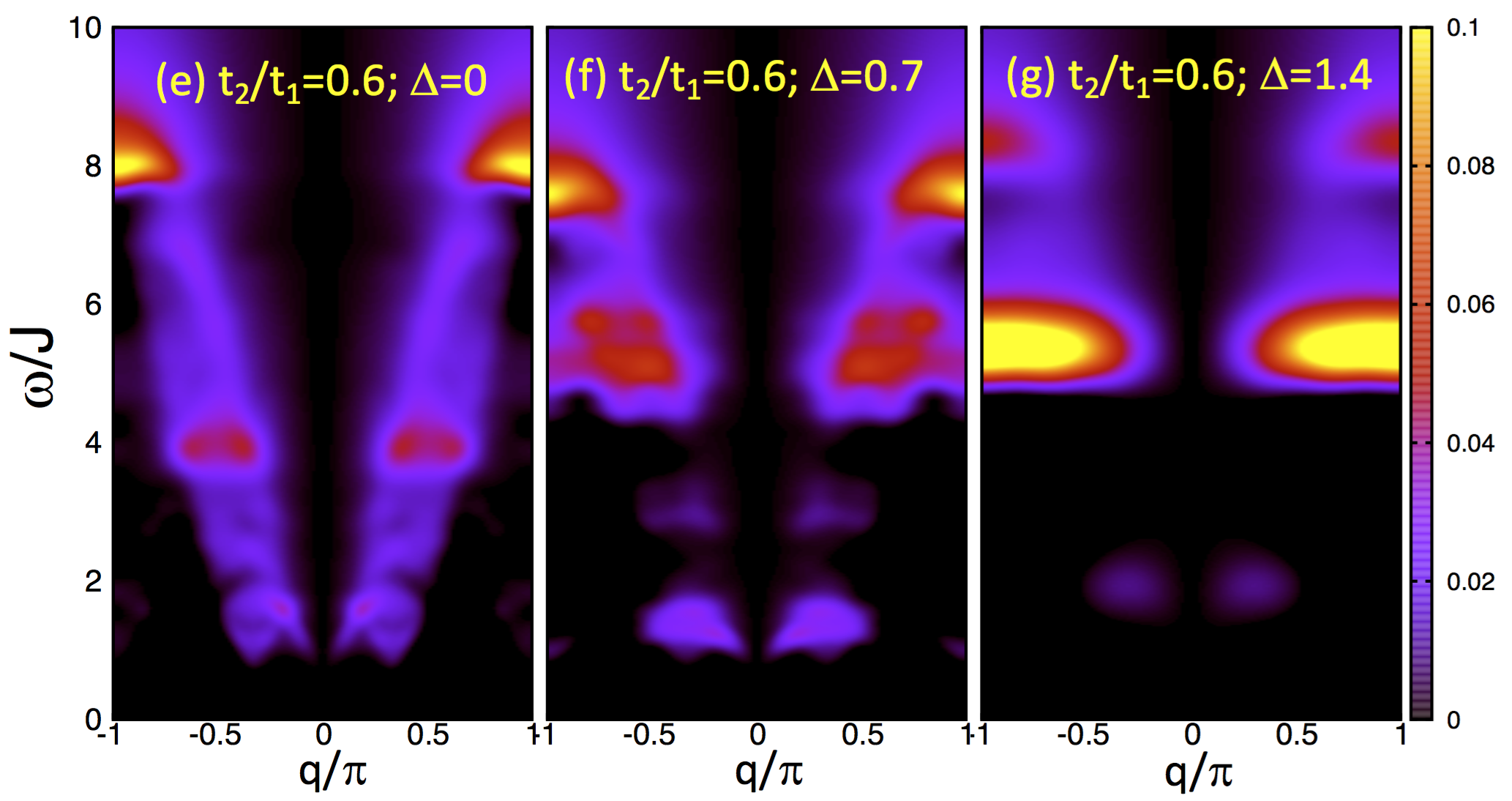}
  \vspace{-0.2cm}
\caption{\label{SU6}Spin spectral function in the folded Brillouin Zone for  for various values of  $t_2/t_1$ and the crystal field. With increase of $\Delta/J$ or the anisotropy the spectral weight shifts toward the dimerized configuration where the singlet-triplet gap is equal to $2J$ corresponding  to the breaking of a dimer. }
\vspace{-0.5cm}
\end{figure}


In reality the SU(6) symmetry is broken which is expressed in the anisotropy of the exchange integrals corresponding to different orbitals and in the presence of the crystal field. The former anisotropy originates from two factors: the difference between tunneling matrix elements of different orbital states and the Hund's coupling. Since the lowest $d$-orbital is occupied by one electron(hole), the Hund's coupling affects only the excited states. 
As shown in Fig.~\ref{fig:relabel}(a), the  strong electron hopping integrals are the head-on $d_{zx}-d_{zx}$ (between the 1st and 2nd Ti atoms) and the head-on $d_{xy}-d_{xy}$ (between the 4th and 5th Ti atoms), whose strength is referred to as $t_1$. All the $t_1$ paths are also depicted as solid arcs in Fig.~\ref{fig:relabel}(b). Yet, for the edge-sharing $t_{2g}$ connections, it is known that the oxygen $p$-orbital-mediated shoulder-to-shoulder hopping paths, e.g., the $d_{zx}-p_{z}-d_{yz}$ between the 2nd and 3rd $M$ atoms in Fig.~\ref{fig:relabel}(a), may be as strong \cite{khomskii,mila,jackeli}. These indirect paths are referred to as $t_2$ and shown as the dashed lines in Fig.~\ref{fig:relabel}(b). Note that the $M$ $yz$ orbitals are involved in the $t_2$ paths only [Fig.~\ref{fig:relabel}(b)]; therefore, in the limit of small $t_2$ or large $t_{2g}$ splitting $\Delta$ (i.e., the $yz$ orbital is higher in energy by $\Delta$ than the $xy$ and $zx$ orbitals), $d_{yz}$ becomes irrelevant, yielding the minimal two-orbital model \cite{brink,brinkEPL,motome}. On the other hand, for considerable $t_2$ and small $\Delta$, the $t_1$ and $t_2$ paths seem to be highly entangled as shown in Fig.~\ref{fig:relabel}(b); however, following the red, blue, and green lines, we found that they can be completely decoupled to form three degenerate hopping paths as shown in Fig.~\ref{fig:relabel}(c). In this sense, the most remarkable property of Na$M$Si$_2$O$_6$ is that its electronic band is exactly 3 times degenerate. In real space the degeneracy is reflected as the following property of the single electron wave functions: $
\psi_b(k+1) = \psi_b(k) = \psi_c(k-1).$

The corresponding band Hamiltonian in notations depicted on Fig.~3(c) has three $M$ sites in the unit cell and is expressed as follows:
\bea
H = -\sum_{k,\alpha=a,b,c} \psi^+_{\alpha,\s}(k)\left(
\begin{array}{ccc}
0 & t_1 & t_2\re^{-3\ri k}\\
t_1 &0 & t_2\\
t_2\re^{3\ri k} & t_2 & \Delta
\end{array}\right ) \psi_{\alpha,\s}(k) \nonumber
\eea
The spectrum is determined by the cubic equation
\be
\epsilon^3 - \epsilon^2\Delta - \epsilon(2t_2^2 +t_1^2) + \Delta t_1^2 - 2t_1t_2^2\cos 3k =0.
\ee
At $t_1=t_2, ~~\Delta =0$ the solution is $\epsilon = 2t\cos k$. The band is 1/6-filled with $k_F = \pi/6$. At $t_2 \neq t_1$ and $\Delta \neq 0$,   spectral gaps appear at $k = \pm \pi/3, \pm 2\pi/3$ corresponding to  the perturbations with wave vectors $q= \pm 2\pi/3,\pm 4\pi/3$. Since they do not coincide with $2k_F$, the weakly interacting electron system would remain gapless \cite{ep}. As we shall see, for the Mott insulator this  is no longer the case. Besides the charge (Mott) gap, which is always present, in the presence of anisotropy our system acquires spectral gaps in all other sectors. This is obviously related to the fact that the perturbations around the SU(6) symmetric point generate relevant operators  with the wave vector $4k_F$.


Integrating over the high energy states we obtain the following Hamiltonian: 
\begin{widetext}
\bea
&& H = \frac{2t_2^2}{U-\Delta}\sum_k P^{o,s}_{k,k+1} + \Delta\sum_k [X_{aa}(3k+2) + X_{bb}(3k)  + X_{cc}(3k+1)] \hat I + \sum_k \delta V_k , \label{H}\\
&& \delta V_k = 2\Big(\frac{t_1^2}{U} - \frac{t_2^2}{U-\Delta}\Big)\times\label{pert}\\
&& \Big[\hat P^s_{3k,3k+1}X_{aa}(3k) X_{aa}(3k+1) + \hat P^s_{3k+1,3k+2}X_{bb}(3k+1)X_{bb}(3k+2) + \hat P^s_{3k+2,3k+3}X_{cc}(3k+2)X_{cc}(3k+3)\Big] + \nonumber\\
&& 2t_2\Big(\frac{t_1}{U} - \frac{t_2}{U-\Delta}\Big)\Big\{\Big[\hat P^s_{3k,3k+1}X_{ab}(3k)X_{ba}(3k+1) +\nonumber\\
&&  \hat P^s_{3k+1,3k+2}X_{ab}(3k+1)X_{ba}(3k+2) + \hat P^s_{3k+2,3k+3}X_{ac}(3k+2)X_{ca}(3k+3)\Big] + H.c.\Big\} + \frac{2t_2(t_1- t_2)}{U-\Delta}\times\nonumber\\
&& \Big\{\Big[\hat P^s_{3k,3k+1}X_{ac}(3k)X_{ca}(3k+1) + \hat P^s_{3k+1,3k+2}X_{bc}(3k+1)X_{cb}(3k+2) + \hat P^s_{3k+2,3k+3}X_{bc}(3k+2)X_{cb}(3k+3)\Big] +H.c.\Big\}, \nonumber
\eea
\end{widetext}
where  $P^s_{k,k+1}$ is the spin permutation operator and $X_{ab}$ are Hubbard operators acting on orbital indices,  defined as $(X_{pq})^{\alpha\beta} = \delta_p^{\alpha}\delta_q^{\beta}$. \red{In \cite{SI} where the derivation is given, this Hamiltonian is written in terms of the isospin operators. Since the Hund's coupling is just affects the anisotropy of the exchange integrals, we set to zero to simplify the calculations. Below we will consider various values of $t_2/t_1$ and $\Delta$.}

\begin{figure}[t]
   \includegraphics[width=0.45 \textwidth]{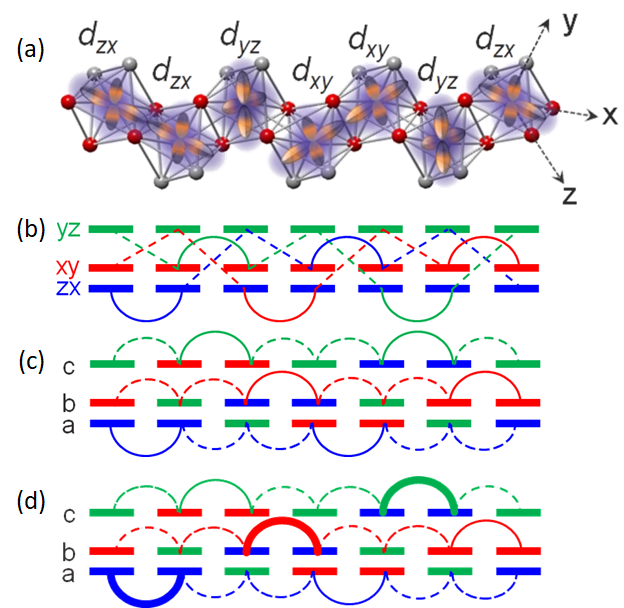}
   \vspace{-0.2cm}
  \caption{\label{fig:relabel}a.) A graphic description of a dimerized state for an isolated {\M}-chain. 
   Only $t_{2g}$ orbital of $M$ ions are depicted. b.) The original tunneling scheme. c.) The tunneling scheme with relabeled orbitals. We relabel  the orbitals on different sites to make the tunneling diagonal. The solid lines correspond to matrix element $t_1$, the dashed lines corresponds to matrix element $t_2$. d.)  The dimerization pattern in the presence of crystal field. The orbitals on which spin singlets form are shown by thick lines. }
\vspace{-0.5cm}
\end{figure}

To get the overall picture of the correlations we used the DMRG method \cite{White2004,Daley2004} to calculate the imaginary part of the correlation function

\bea
S(\omega, q) = \sum_k \int_0^{\infty} \rd t \la S_k^z(t)S_m^z(0)\ra\re^{\ri\omega t + \ri q(k-m)},
 \eea
where $S^z_k$ is the spin projection operator acting on site $k$.  The spectral weight contains rich information about the excitation spectrum of the model.
We carry out calculations with a Suzuki-Trotter decomposition of the evolution operator \cite{vietri,Paeckel2019} and a time-step $\delta t=0.1$ in units of $1/J$. We have been able to study chains with up to 48 unit cells ($L=144$ sites) using up to 1600 DMRG states for the time evolution, and 5000 for ground state calculations, that translates into a truncation error of $10^{-5}$ and $10^{-8}$ respectively for the gapless case (similar accuracy is obtained in the gapped case with a smaller basis size). Most time-dependent simulations were conducted on chains with  $24$ unit cells ($L=72$ sites). The local space of configurations has dimension 6, but we use $U(1)$ symmetry corresponding to $S^z$ and density conservation for each orbital channel (4 quantum numbers in total). The density for each orbital sector is fixed at $n=1/3$, while the spin is set to $S^z=0$. This is equivalent to density $n=1/6$ in the $SU(6)$ chain\cite{Manmana2011}.
We calculate the spectral function in real time and space with open boundary conditions, and Fourier transform it to obtain resolution in momentum and frequency following the prescription outlined in Refs.\onlinecite{White2008,vietri,Paeckel2019}.




\emph{Limit of small $\Delta/J$, $t_2/t_1 =1$.---} Having in mind a broader aim than a particular case of  {\Na}, we deem it instructive to set the SU(6)-invariant model as its starting point of our analysis.  The SU(6)-symmetric limit $\Delta =0, t_2 = t_1$ allows an analytical treatment. The thermodynamics and the excitation spectrum are extracted from Bethe ansatz. At low energies the spectral function can  be analyzed by means of Conformal Field Theory. At higher energies one can also use the $1/N$-expansion.

As expected from the exact solution, the spectrum of the SU(6) symmetric model is gapless and the main spectral weight is centered at $q = \pm \pi/3$ which corresponds to $\pm 2k_F$.
The spectral function  also looks squeezed into the region
\be
4J\sin(q/2)\sin|k_F- q/2| < \omega < 4J\sin(q/2)
\ee
corresponding to two-spinon emission. This agrees very well with $1/N$ picture where the spinons are represented as weakly interacting fermions and the spin operator is quadratic in the fermions.  In the presence of anisotropy spectral gaps open at $q = \pm 2k_F = \pm \pi/3$ shown on Fig. 2 meaning that the anisotropy generates a relevant operator which carries momentum $4k_F$. Such operator is indeed present at the SU(6) Quantum Critical Point, it transforms according to the representation of the SU(6) group with the Young tableau consisting of a vertical column with two boxes. The scaling dimension is $d = 2(1-1/N) = 5/3$. The presence of such perturbation as one might expect, also leads to spontaneous dimerization (see Figs.~\ref{fig:relabel}(a)(d) and Fig. \ref{Dimerization}). This order breaks a discrete (translational) symmetry, all other fluctuations are gapped and short range.  Obviously, small perturbations preserve the SU(6) structure of the particle multiplets such that spin and orbital excitations are degenerate. The spectral gaps  grow slowly with $\Delta/J$ as shown on Fig. (\ref{gap})  due the high value of the scaling dimension of the perturbing operator. Hence the SU(6) symmetry is preserved at low energies: Fig. (\ref{gap}) shows that at $\Delta/J <0.4$ a difference between the gaps for excitations with different quantum numbers is practically undetectable. At larger anisotropies the multiplets will be split. 

\begin{figure}[t]
   \includegraphics[width=0.45 \textwidth]{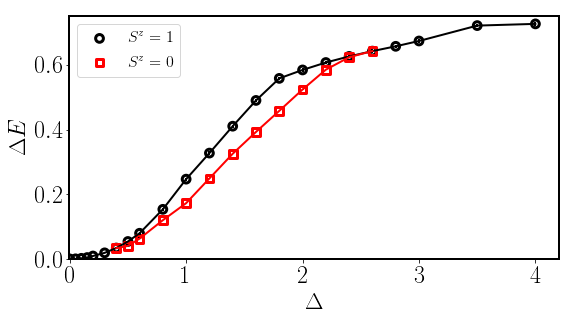}
   \vspace{-0.2cm}
  \caption{\label{gap}Numerical results for the lowest spectral gaps for various values of the crystal field $\Delta$ and $t_2 = t_1$.}

\vspace{-0.5cm}

\end{figure}

\begin{figure}[t]
   \includegraphics[width=0.45 \textwidth]{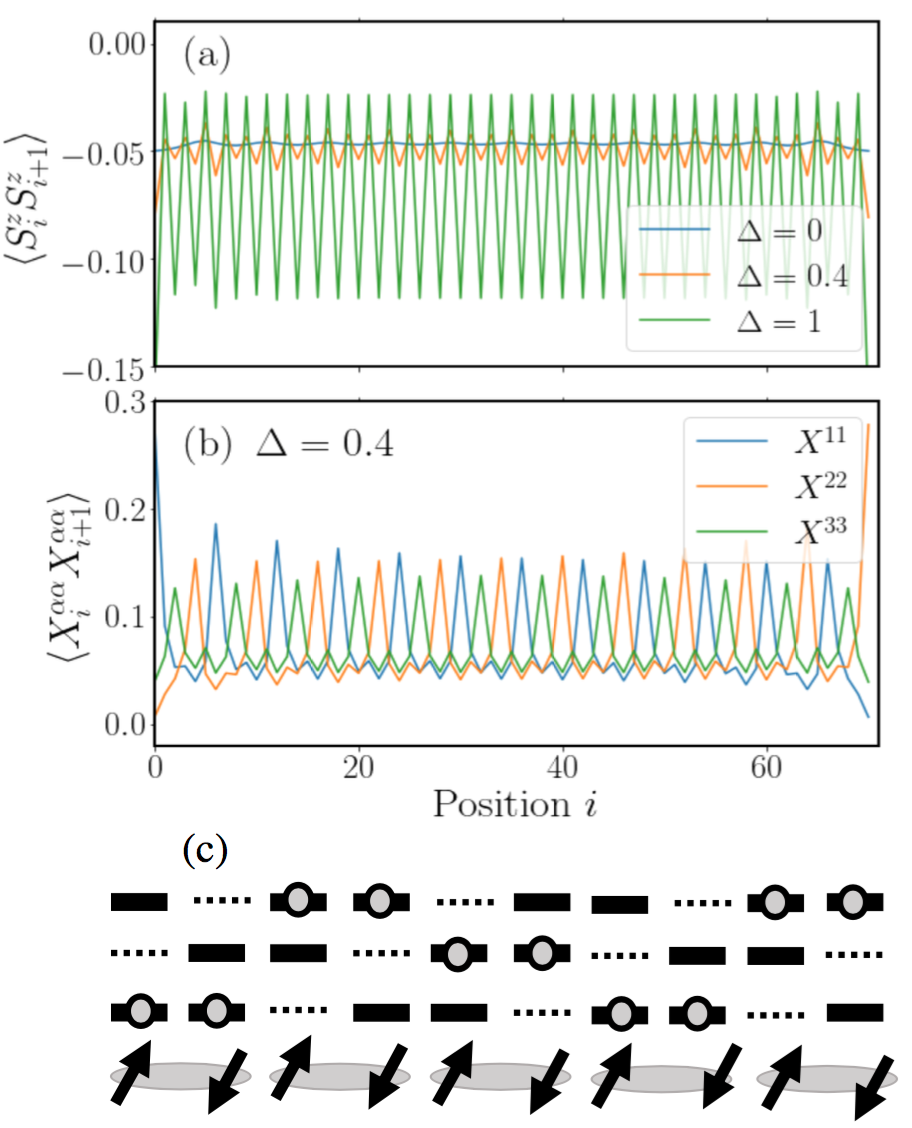}
   \vspace{-0.2cm}
  \caption{\label{Dimerization} Dimerization for various values of the
crystal field $\Delta$ and $t_2 =t_1$.  (a) Nearest neighbor spin-spin
correlation; (b) density-density correlation in the orbital channel; (c)
schematic illustration of the spin-orbital order in the limit of large
$\Delta$: dashed lines represent orbitals that are projected out. Charge
fluctuations are suppressed and charge is frozen in the depicted
pattern. Spin is only allowed to interact in pairs forming independent
singlets. }

\vspace{-0.5cm}

\end{figure}

\emph{Limit of large $\Delta/J$.---}The easiest way to understand the dimerization phenomenon is to consider the limit of large crystal field. For $J=0$ each site has two degenerate orbitals in the ground state. For sites $3n$ it may be (1,2), for $3n+1$ - (1,3), for $3n+2$ - (2,3), {\it etc.}. At $J\neq 0$ the degeneracy is lifted and the ground state becomes dimerized. One possible sequence of occupied orbitals is $(1,1,2,2,3,3, ...)$ which corresponds to nonvanishing exchange between sites (3n,3n+1), (3n+2,3n+3), {\it etc.}(see Fig ~\ref{fig:relabel}(d)). The other sequence is $(2,3,3,1,1,..)$ with nonvanishing exchange between (3n+1,3n+2), (3n+3,3n+4), {\it etc.} So, in the limit of infinite $\Delta$ the ground state consists of isolated periodically arranged spin dimers. Our numerical calculations demonstrate that the dimerization persists down to smallest values of $\Delta/J$ (see Fig.\ref{Dimerization}). As far as the spectrum is concerned, it leads to two major effects. First, it opens gaps for all excitations. Second, it leads to a progressive shift of the spectral weight towards frequency $\omega = 2J$ corresponding to the breaking of an isolated dimer \red{[see Figs. \ref{SU6}(c)(d)]}. Nevertheless, there is some weight at about $J/2\sim 13$ meV, given $2J\simeq 53$ meV \cite{spingap}, in agreement with the gap seen in the heat capacity data \cite{JPSJ}.


According to the first-principles calculations and Wannier function analysis \cite{SI,yin_prb09}, {\Na} has the following parameters:
$U = 3.8$ eV, $J_H = 0.8$ eV, $t_1 =0.203$ eV, $t_2/t_1 = 0.21$, $t_1^2/U \approx 0.01$ eV. Hence in {\Na} the deviation from $t_2/t_1=1$ is quite significant. However, due to the well-known double counting issue on the LDA+$U$ approach to correlated materials, the value of $\Delta$ is uncertain and is taken as a free parameter. As we have seen at moderate values of anisotropy and crystal field the excitations are gapped and become practically dispersionless (Figs.\ref{SU6}(d,f,g)) corresponding to the local dimers discussed above. This, in all likelihood, is the situation in {\Na} which thus fails our expectations for an orbital spin liquid. However, as follows from Figs. \ref{SU6}(b,c), at moderate values of the anisotropy and crystal field there is a significant spectral weight at small energies. The spectral function bears some resemblance to the SU(6)-symmetric one which a is sign that the orbital degrees of freedom are not quenched. We suggest that such situation may exist in the ruthenium- or osmium-based pyroxenes where Ru$^{3+}$ or Os$^{3+}$ ions contain one $t_{2g}$ hole. These are candidates for liquids with tightly bound spin and orbital excitations. In the early $3d$ transition-metal oxides such as the titanium oxide, the $3d$ energy is considerably different from the oxygen $p$ orbitals, which creates the barrier that hinders the indirect hopping $t_2$. However, $t_2$ may become dominant as in, for example, Na$_2$IrO$_3$ and RuCl$_3$ to induce the Kitaev-type spin frustration \cite{jackeli}. 
Specifically, considering Ru$^{3+}$ has almost the same Shannon ionic radii as Ti$^{3+}$, we did similar first-principles calculations for NaRuSi$_2$O$_6$ (Supplemental Material \cite{SI}). We found that $t_2/t_1 = 0.64$ ($t_1 =0.132$ eV), which is much more favorable than the {\Na} case. In addition, the $yz$ orbital moves higher in energy, which however is closer to and mixed with the hole bands of the $xy$ and $zx$ characters. Moreover, the experimentally observed large bond dimerization is favored in the first-principles calculation for {\Na} but not for NaRuSi$_2$O$_6$. Thus, it would be interesting to synthesize NaRuSi$_2$O$_6$ and compare its low-energy physical properties with the present theory.


This work was supported by the Office of Basic Energy Sciences, Material Sciences and Engineering Division, U.S. Department of Energy (DOE) under Contract No. DE-SC0012704 (A.M.T., W.Y. and E.S. B.) and DE-SC0014407 (A.E.F.).


\section{[Supplemental Material] Quantum liquid with strong orbital fluctuations: The case of pyroxene family}

\subsection{First-principles calculations}
First-principles band structure calculations were performed by using the WIEN2K \cite{wien2k} implementation of the full potential linearized augmented plane wave method in the generalized gradient approximation (GGA) \cite{gga} of DFT. For the high-temperature structure (HTS; space group C2/c) and low-temperature structure (LTS; space group P$\bar{1}$) of {\Na}, we used the x-ray diffraction data at $T=298$ K and $T=100$ K, respectively. To make direct comparison, we managed to apply the P$\bar{1}$) space group to HTS, too. The Ti zigzag chains run along the crystallographical $a$ direction. The basis size was determined by $R_\mathrm{mt}K_\mathrm{max}$ = 7 and the Brillouin zone was sampled with a regular $13\times 10 \times 10$ mesh containing 196 irreducible $k$ points to achieve energy convergence of 1 meV. Electron hopping integrals were obtained by using Wannier function analysis of the GGA band structures. \cite{ku_prl02,yin_prl06,yin_prl06_TMD,yin_prb09}.

For {\Na}, Fig.~\ref{fig:BS} shows that the $e_g$ orbitals stay about 2 eV higher than the $t_{2g}$ orbitals. For nonmagnetic cases, the total energy of LTS is lower by 29 meV/f.u. than that of HTS. That is, LTS is more stable than HTS at $T=0$, in agreement with experiments. In Table~\ref{table:hopping}, we present the hopping parameters between $M$ atoms: $t_1=-0.2025$ eV, $t_2=-0.0427$ eV, and $t_2/t_1=0.21$.

\red{As for {\M} where $M=$Ru, we have not found any report on its structural data. Considering Ru$^{3+}$ has almost the same Shannon ionic radii as Ti$^{3+}$, we did similar first-principles calculations using the structural data of {\Na}. Fig.~\ref{fig:Ru} shows that the Ti and Ru cases have one $t_{2g}$ electron and hole on each $M$ site, respectively, and that the $yz$ orbital is mixed into the $xy/zx$ hole bands in the Ru case more substantially than into the $xy/zx$ electron bands in the Ti case. Table~\ref{table:hopping} shows that $t_1=-0.1323$ eV, $t_2=-0.0849$ eV, and $t_2/t_1=0.64$.  At a glance, the $yz$ orbital moves higher in energy, which however is closer to and mixed with the hole bands of the $xy$ and $zx$ characters. This, together with the tripled $t_2/t_1$, makes the total energy of LTS higher than that of HTS by 29 meV/f.u., indicating less tendency to dimerization in {\Ru}.}

\begin{table*}[t]
\caption{\label{table:hopping}%
\red{Nearest-neighbor hopping parameters between $M$ atoms obtained from the Wannier functions analysis of the GGA electronic
structures of {\M} where $M$ = Ti and Ru. The unit is eV.}} %
\begin{ruledtabular}
\begin{tabular}{r|rdddcddd}
  &\multicolumn{3}{c}{\Na}& & \multicolumn{3}{c}{\Ru}\\
     &    $yz$  &  $zx$   &   $xy$  &  & $yz$  &  $zx$   &   $xy$ \\
  \hline
$yz$ &  0.0838  &  0.0204 &  -0.0427 & &  0.0641  &   0.0142  &  -0.0849\\
$zx$ &  0.0204  & -0.2025 &  -0.0182 & &  0.0142  &  -0.1323  &  -0.0438\\ 
$xy$ & -0.0427  & -0.0182 &   0.0875 & & -0.0849  &  -0.0438  &   0.0629\\ 
 \hline
\end{tabular}
\end{ruledtabular}
\end{table*}

\begin{figure}[t]
\includegraphics[width=\columnwidth,clip=true,angle=0]{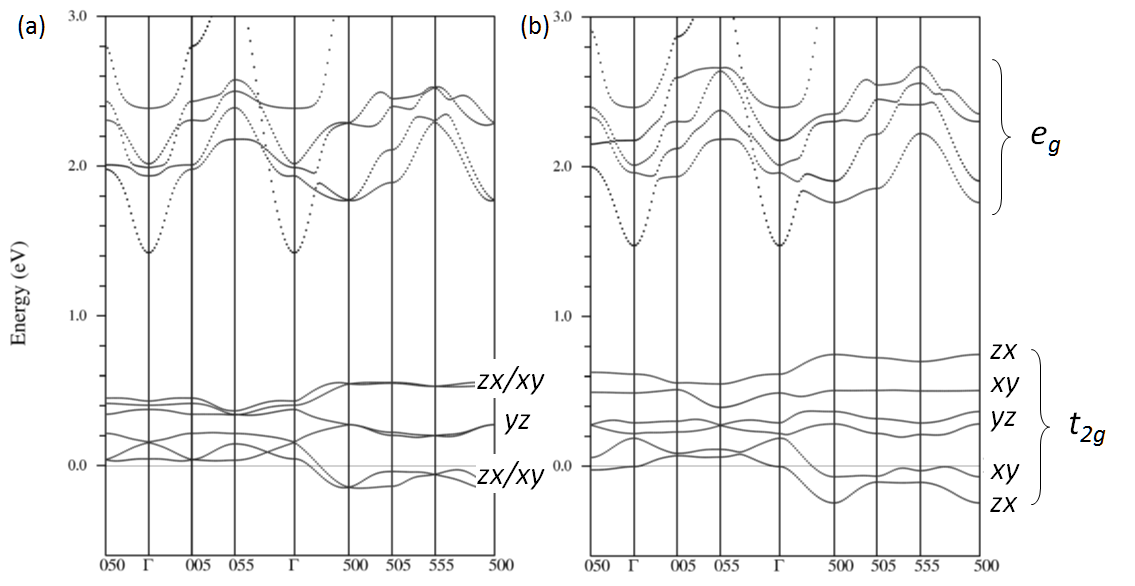}
\caption{\label{fig:BS}Band structures of nonmagnetic {\Na} for HTS (a) and LTS (b). The $x$-axis labels 005, 055, 500, 505, 555 mean the electronic momentums at $(0,0,\pi)$, $(0,\pi,\pi)$, $(\pi,0,0)$, $(\pi,0,\pi)$, $(\pi,\pi,\pi)$, respectively. The orbital characters at $(\pi,0,0)$ are shown.}
\end{figure}

\begin{figure}[t]
\includegraphics[width=\columnwidth,clip=true,angle=0]{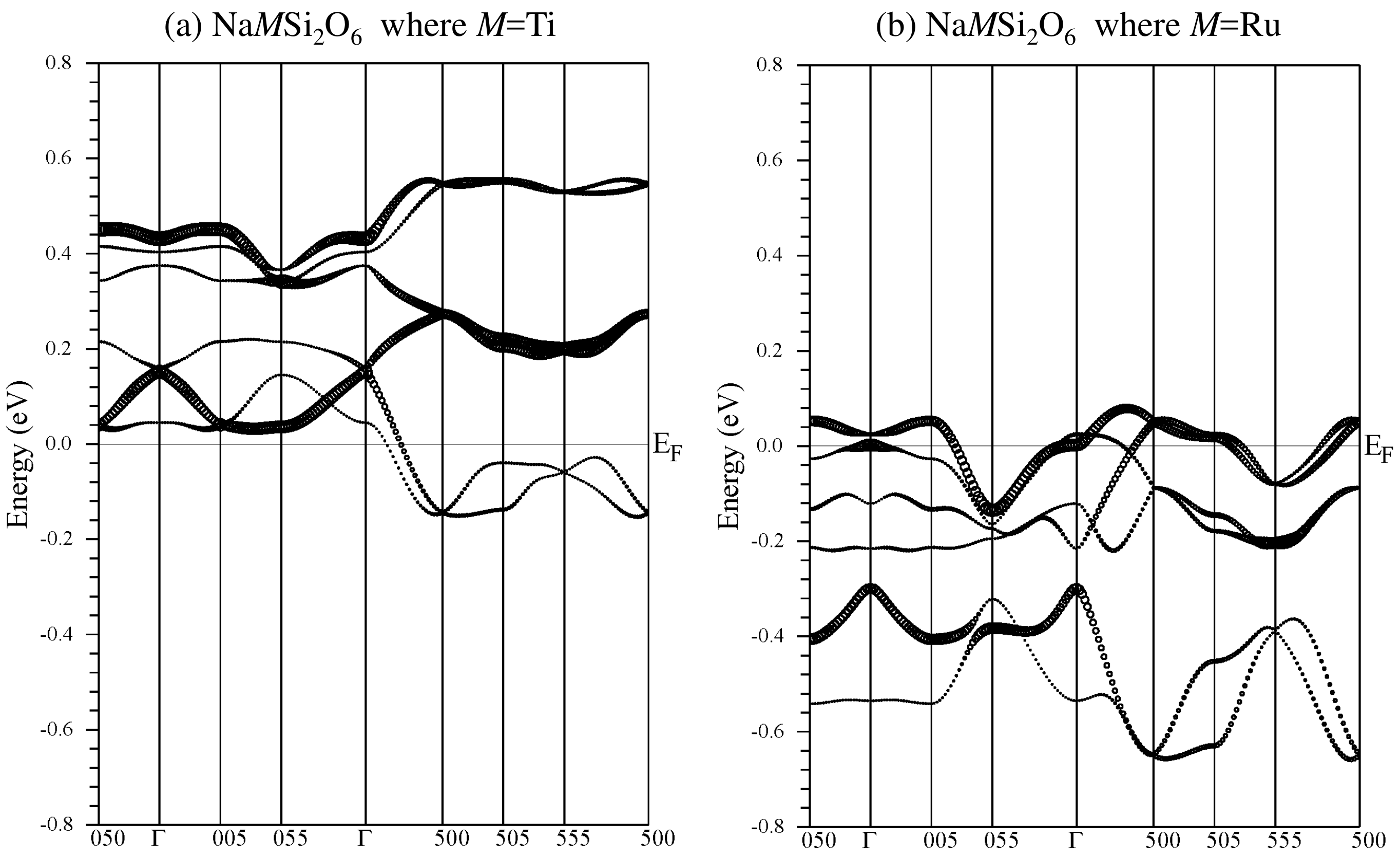}
\caption{\label{fig:Ru}\red{Band structures of {\M} with nonmagnetic HTS for (a) $M=$Ti and (b) $M=$Ru. The weight of the $yz$ orbital characters is shown by the size of the circles.}}
\end{figure}

\subsection{{\Na} is a Mott insulator}
In GGA, the system is metallic and becomes insulating in the GGA+$U$ calculations. The band gap size appears to be linearly proportional to $U$ [Fig.~\ref{fig:TotalE}(a)], while the Ti magnetic moment is less sensitive to $U$ for $U\geq 1$ eV [Fig.~\ref{fig:TotalE}(b)]. The bandwidth of the $t_{2g}$ bands are all very narrow (Fig.~\ref{fig:BS_AF_U3} and Fig.~\ref{fig:BS_FM_U3}),
These behaviors hold for both the AF and FM cases and both HTS and LTS, indicating that {\Na} is a Mott insulator. The realistic $U$ is about $3-4$ eV to reproduce the gap size that corresponds to the greenish appearance of {\Na}.

\begin{figure}[h]
\includegraphics[width=\columnwidth,clip=true,angle=0]{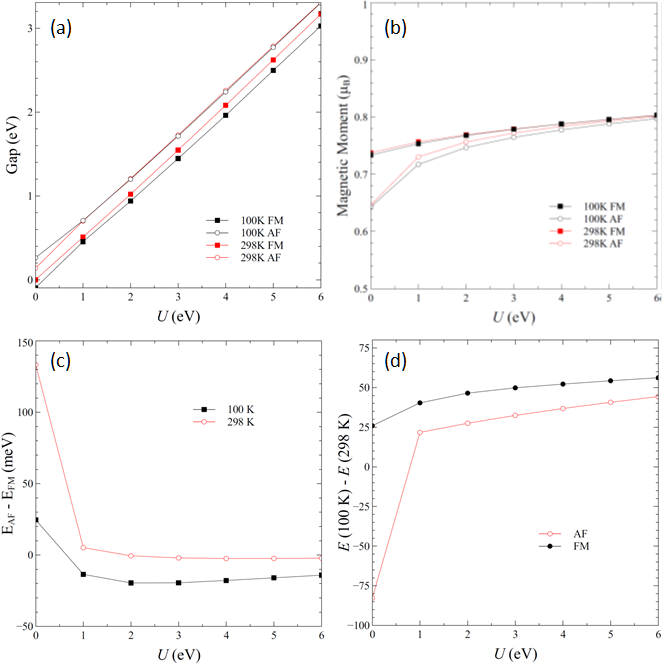}
\caption{\label{fig:TotalE}
The $U$ dependence of (a) Band gap size, (b) Ti magnetic momentum, (c) total energy difference between the AF and FM states, and (d) total energy difference between LTS and HTS.}
\end{figure}

\begin{figure}[h]
\includegraphics[width=\columnwidth,clip=true,angle=0]{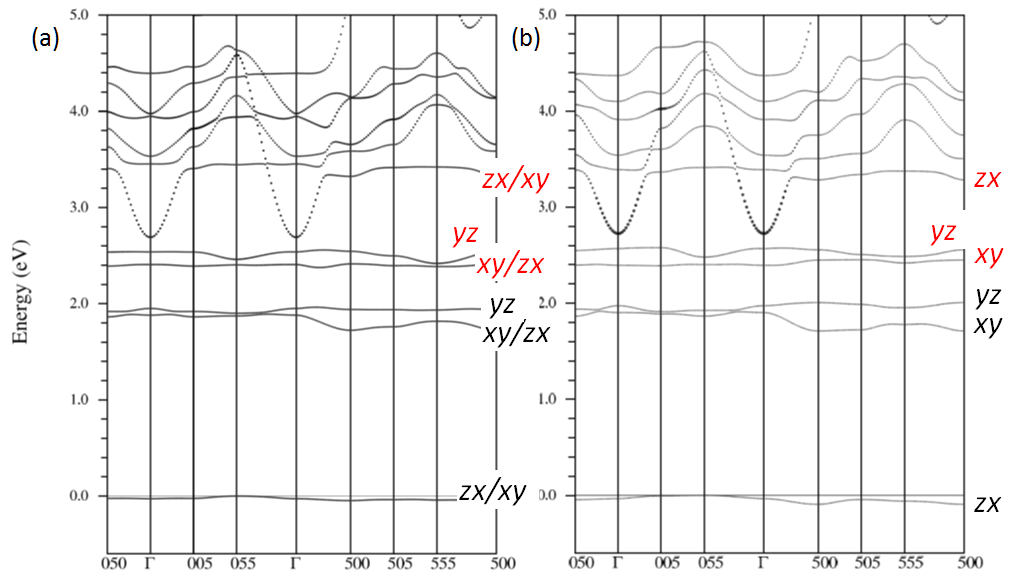}
\caption{\label{fig:BS_AF_U3}Band structures of AF {\Na} for (a) HTS and (b) LTS. $U=3$ eV.}
\end{figure}

\begin{figure}[h]
\includegraphics[width=\columnwidth,clip=true,angle=0]{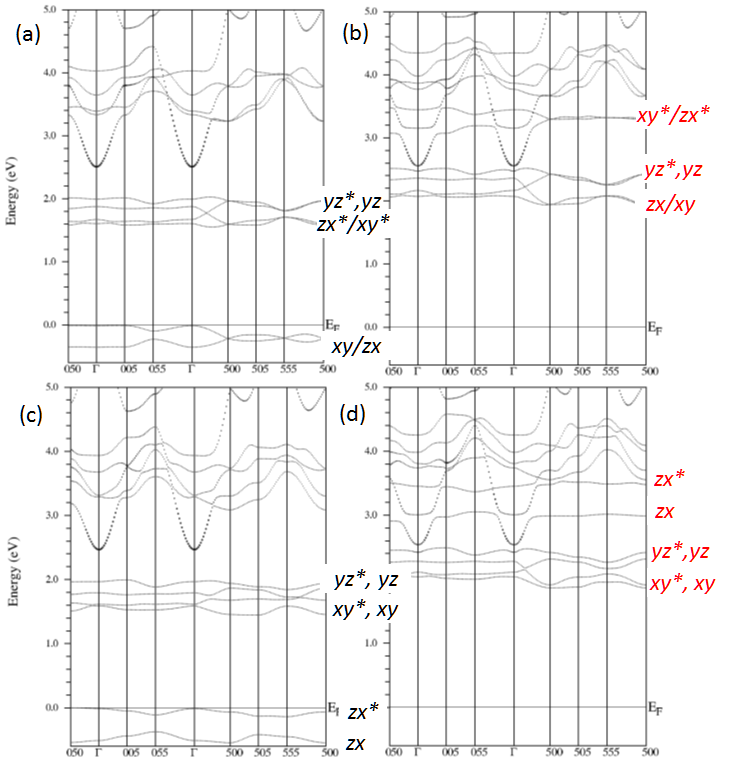}
\caption{\label{fig:BS_FM_U3}Band structures of FM {\Na} for spin majority in (a) HTS and (b) LTS and for spin minority in (c) HTS and (d) LTS . $U=3$ eV.}
\end{figure}

As shown in Fig.~\ref{fig:TotalE}(c), for LTS, AF is stabler by about 20 meV/f.u. than FM for $U\geq 1$ eV, a favorable condition for spin-singlet formation. For HTS, AF and FM are almost degenerate; hence, the magnetic state in HTS is actually paramagnetic.

However, as shown in Fig.~\ref{fig:TotalE}(d), HTS is considerably lower in energy by $20-40$ meV/f.u. than LTS. This unsatisfactory result implies that strong quantum spin fluctuations are crucial to stabilize LTS in reality.
\ignore{This could be understood in the minimal electronic model for the two-orbitally assisted Peierls transition, which is given by
\begin{equation}
H=J\sum_{\langle i,j\rangle}{\mathbf{S}_i\cdot\mathbf{S}_j\Big[\frac{1}{4}+T^z_iT^z_j+\frac{(-1)^i}{2}(T^z_i+T^z_j) \Big]}.
\end{equation}
In HTS, one may take the mean-field values $\langle T^z_i \rangle=\langle T^z_j \rangle=\langle T^z_iT^z_j \rangle=0$. In LTS, $H=J\sum_{\langle i,j\rangle}{\mathbf{S}_i\cdot\mathbf{S}_j}$ on the short bonds and vanishes on the longer bonds. In the classical spin limit, the bond energy of the $S=\frac{1}{2}$ is $-JS^2=-J/4$. The AF-FM energy difference 18meV/f.u. -> 36meV/short bond -> J=72 meV. By contrast, the bond energy of a quantum spin singlet is $-3J/4$; $J/2 /short bond$ lower than classic, the additional average bond energy gain in LTS is thus $J/4=18 meV$.
}

\subsection{Effective low-energy model Hamiltonian}

\begin{figure}[t]
\vspace{-0.5cm}
        \includegraphics[width=0.8\columnwidth,clip=true,angle=0]{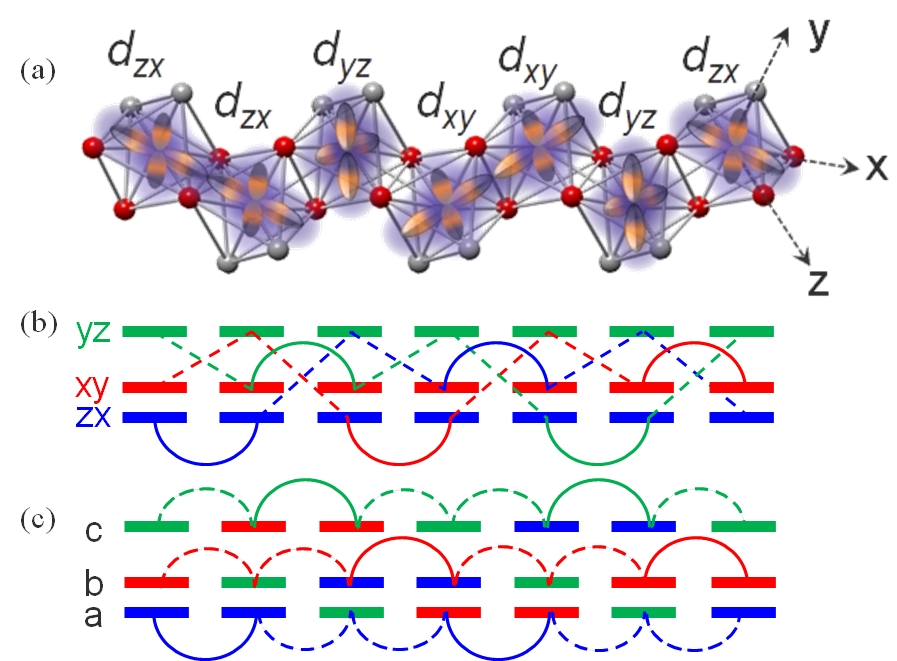}
        \caption{\label{crystal}
        (a) Crystal structure of the Ti-O chain in NaTiSi$_2$O$_6$. (b) Electron hopping paths with respect to fixed coordinate axes. (c) Electron hopping paths with respect to locally rotated coordinate axes. Solid and dashed lines denote the $t_1$ and $t_2$ paths, respectively.
        \label{geometry}}
\end{figure}

The three Ti $3d$ $t_{2g}$ orbitals ($d_{xy}$, $d_{yz}$, $d_{zx}$) form a practically complete basis for the low-energy Hilbert space. A general 3-orbital Hamiltonian $H$ is given by
\begin{eqnarray}%
H&=& -\sum_{ijmm^\prime\sigma} {(t^{mm^\prime}_{ij}
C_{im\sigma}^\dag C_{jm^\prime\sigma}^{} + H.c.)} + \sum_{im\sigma}
{\varepsilon_{m} n^{}_{im\sigma}} \nonumber\\ &+& U_\mathrm{eff}
\sum_{i,m=1,2} {n_{im\uparrow} n_{im\downarrow } } +
U^\prime_\mathrm{eff} \sum_{i\sigma\sigma^\prime} { n_{i1\sigma}
n_{i2\sigma^\prime} } \nonumber \\ &-& J_\mathrm{eff} \sum_{i\sigma\sigma^\prime}
{C_{i1\sigma}^\dag C_{i1\sigma^\prime}^{} C_{i2\sigma^\prime}^\dag
C_{i2\sigma}}^{}, \label{eq:model}
\end{eqnarray}
\noindent where $C_{im\sigma}$ annihilates an electron with spin
$\sigma$ in the $m$-th orbital at site $i$. Here $m=1,2,3$ stand for Ti $d_{xy}$, $d_{yz}$, $d_{zx}$ orbitals, respectively. the $U$ and $U'$ terms are the intraorbital and interorbital Coulomb interactions, respectively, and $J$ is the strength of Hund's rule coupling. In the ideal case of no distortion nor tilting of the edge-sharing TiO$_6$ octahedra in NaTiSi$_2$O$_6$, the leading nearest-neighbor hopping parameters satisfy [Fig.~\ref{fig:t1-t2}]
\begin{eqnarray}
t^{11}_{2m,2m-1}&=&t_1,\nonumber \\
t^{23}_{2m,2m-1}&=&t_2,\nonumber \\
t^{32}_{2m,2m-1}&=&t_2,\nonumber \\
t^{22}_{2m,2m+1}&=&t_1,\nonumber \\
t^{13}_{2m,2m+1}&=&t_2,\nonumber \\
t^{31}_{2m,2m+1}&=&t_2,\nonumber \\
\textrm{and 0 for} &\text{the} & \text{others}.
\end{eqnarray}
There are orbital-crossing hopping parameters [Fig.~\ref{crystal}(b)]. However, the crossing terms will disappear in the relabelling of the local orbitals as shown in Fig.~\ref{crystal}(c). This turns the number of Ti atoms in the unit cell from two to six. The resulting $a$, $b$, $c$ orbitals are decoupled in the kinetic-energy terms. They are coupled by the interaction terms $U$, $U'$, and $J$ in the same way as defined in Eq.~(\ref{eq:model}).

For $U$, $U'$, and $J >> t^{mm'}_{ij}$, the states with two or more electrons occupying the same Ti site can be projected out by second-order perturbation or the following canonical transformation \cite{yin_prb09}:
Suppose $H=H_0+H_1$, and define $H(\lambda)=H_0 + \lambda H_1$. Then, the transformed Hamiltonian is
\begin{eqnarray}
H_S(\lambda)&=&e^{-\lambda S}He^{\lambda S} \simeq H_0+
\frac{\lambda^2}{2!}[ H_1, S],
\end{eqnarray}
after removing the terms linear with respect to $\lambda$ following
\begin{equation}
    \lambda H_1 + [H_0,\lambda S] = 0. \label {eq:S}
\end{equation}

Let us apply the above canonical transformation to
Eq.~(\ref{eq:model}). First define the Hubbard operator on the
$i$-th site as
\begin{equation}
X_i^{pq}=|p\rangle^{}_i\langle q|^{}_i,
\end{equation}
and the projection operator as
\begin{equation}
P_0=\prod_i\sum_{p \in g.s.} X_i^{pp}, \;\;\;\; P_1 = 1-P_0,
\end{equation}
where g.s. means the ground state multiplets (the states we want to
keep); here they are empty or singly occupied states. Then,
$H_0$ and $H_1$ may be defined by
\begin{eqnarray}
H_0&=&P_0HP_0+P_1HP_1 \nonumber\\
 &=&\sum_i\sum_{p}
{\epsilon_{ip}X_i^{pp}} \nonumber \\
&+& P_0\sum_{i<j}\sum_{rr^\prime ss^\prime}{(V_{ij}^{rr^\prime,
ss^\prime}X_i^{rr^\prime }X_j^{ss^\prime} + H.c.)}P_0 \nonumber \\
&+& P_1\sum_{i<j}\sum_{rr^\prime ss^\prime}{(V_{ij}^{rr^\prime,
ss^\prime}X_i^{rr^\prime }X_j^{ss^\prime} + H.c.)}P_1, \label{eq:H0}
\end{eqnarray}
and
\begin{eqnarray}
H_1&=&P_0HP_1 + P_1HP_0 \nonumber \\
&=& P_0 \sum_{i<j}\sum_{rr^\prime ss^\prime}{(V_{ij}^{rr^\prime,
ss^\prime}X_i^{rr^\prime }X_j^{ss^\prime} + H.c.)} P_1 \nonumber \\
&+& P_1 \sum_{i<j}\sum_{rr^\prime ss^\prime}{(V_{ij}^{rr^\prime,
ss^\prime}X_i^{rr^\prime }X_j^{ss^\prime} + H.c.)} P_0,
\end{eqnarray}
From Eq. (\ref{eq:S}) on the condition we neglect the intersite
terms,
\begin{eqnarray}
S &\simeq& P_0 \sum_{i<j}\sum_{rr^\prime
ss^\prime}{(A_{ij}^{rr^\prime,
ss^\prime}X_i^{rr^\prime }X_j^{ss^\prime} - H.c.)} P_1 \nonumber \\
&+& P_1 \sum_{i<j}\sum_{rr^\prime ss^\prime}{(A_{ij}^{rr^\prime,
ss^\prime}X_i^{rr^\prime }X_j^{ss^\prime} - H.c.)} P_0,
\end{eqnarray}
where
\begin{equation}
A_{ij}^{rr^\prime, ss^\prime}=\frac{V_{ij}^{rr^\prime,
ss^\prime}}{\epsilon_{r^\prime}+\epsilon_{s^\prime}-\epsilon_{r}-\epsilon_{s}}.
\end{equation}

For simplicity, we present the derivation only for the two orbitals with $t^{aa}_{ij}$ and $t^{cc}_{ij}$. The others are easily obtained by replacing the pair of \{$a$, $c$\} with the pair of \{$a$, $b$\} or \{$b$, $c$\}.

We define the spin operator $\mathbf{S}^{}_i$ and the 2-orbital operator $\mathbf{T}^{}_i$ below.
\begin{eqnarray}
\mathbf{S}^{}_i&=&\frac{1}{2}\sum_{m=1,2,3}\sum_{\alpha=\uparrow,\downarrow}\sum_{\beta=\uparrow,\downarrow}
{C_{im\alpha}^\dag \mathbf{\sigma}_{\alpha\beta} C_{im\beta}^{}}, \\  \mathbf{T}^{}_i&=&\frac{1}{2}\sum_{\sigma==\uparrow,\downarrow}\sum_{m=a,c}\sum_{m'=a,c}{C_{im\sigma}^\dag \mathbf{\sigma}_{mm'} C_{im'\sigma}^{}}.
\end{eqnarray}
That is,
\begin{eqnarray}
S^z_i&=&\frac{1}{2}(n_{ia\uparrow}^{}+n_{ib\uparrow}^{}+n_{ic\uparrow}^{}-n_{ia\downarrow}^{}-n_{ib\downarrow}^{}-n_{ic\downarrow}^{}),\nonumber\\
S^+_i&=&C_{ia\uparrow}^\dag C_{ia\downarrow}^{}+C_{ib\uparrow}^\dag C_{ib\downarrow}^{}+C_{ic\uparrow}^\dag C_{ic\downarrow}^{},\nonumber\\
S^-_i&=&C_{ia\downarrow}^\dag C_{ia\uparrow}^{}+C_{ib\downarrow}^\dag C_{ib\uparrow}^{}+C_{ic\downarrow}^\dag C_{ic\uparrow}^{},\\
T^z_i&=&\frac{1}{2}(n_{ia\uparrow}^{}+n_{ia\downarrow}^{}-n_{ic\uparrow}^{}-n_{ic\downarrow}^{}),\nonumber\\
T^+_i&=&C_{ia\uparrow}^\dag C_{ic\uparrow}^{}+C_{ia\downarrow}^\dag C_{ic\downarrow}^{},\nonumber\\
T^-_i&=&C_{ic\uparrow}^\dag C_{ia\uparrow}^{}+C_{ic\downarrow}^\dag C_{ia\downarrow}^{}.
\end{eqnarray}
In terms of the Hubbard operators,
\begin{eqnarray}
S^z_i&=&\frac{1}{2}(X_i^{aa}+X_i^{bb}+X_i^{cc}-X_i^{\bar{a}\bar{a}}-X_i^{\bar{b}\bar{b}}-X_i^{\bar{c}\bar{c}}),\nonumber\\
S^+_i&=&X_i^{a\bar{a}}+X_i^{b\bar{b}}+X_i^{c\bar{c}},\nonumber\\
S^-_i&=&X_i^{\bar{a}a}+X_i^{\bar{b}b}+X_i^{\bar{c}c},\\
T^z_i&=&\frac{1}{2}(X_i^{aa}+X_i^{\bar{a}\bar{a}}-X_i^{cc}-X_i^{\bar{c}\bar{c}}),\nonumber\\
T^+_i&=&X_i^{ac}+X_i^{\bar{a}\bar{c}},\nonumber\\
T^-_i&=&X_i^{ca}+X_i^{\bar{c}\bar{a}},
\end{eqnarray}
where for shorthand notation, $a$ and $\bar{a}$ denote the \emph{singly occupied} spin-up and spin-down orbital-$a$ states, respectively, and likewise for $b$, $\bar{b}$, $c$, $\bar{c}$.

Next, let us work out some math:
\begin{widetext}
\begin{eqnarray}
4 (\frac{1}{4}n_i n_j+S^z_iS^z_j) (\frac{1}{4}n_i n_j-T^z_iT^z_j) &=& X_i^{aa}X_j^{cc}+X_i^{\bar{a}\bar{a}}X_j^{\bar{c}\bar{c}}+X_i^{cc}X_j^{aa}+X_i^{\bar{c}\bar{c}}X_j^{\bar{a}\bar{a}},\nonumber\\
4 (\frac{1}{4}n_i n_j-S^z_iS^z_j) (\frac{1}{4}n_i n_j-T^z_iT^z_j) &=& X_i^{aa}X_j^{\bar{c}\bar{c}}+X_i^{\bar{a}\bar{a}}X_j^{cc}+X_i^{cc}X_j^{\bar{a}\bar{a}}+X_i^{\bar{c}\bar{c}}X_j^{aa},\nonumber\\
2 (S^+_iS^-_j + S^-_iS^+_j) (\frac{1}{4}n_i n_j-T^z_iT^z_j) &=& X_i^{a\bar{a}}X_j^{\bar{c}c}+X_i^{\bar{a}a}X_j^{c\bar{c}}+X_i^{c\bar{c}}X_j^{\bar{a}a}+X_i^{\bar{c}c}X_j^{a\bar{a}},\nonumber\\
2 (\frac{1}{4}n_i n_j+S^z_iS^z_j) (T^+_iT^-_j + T^-_iT^+_j) &=& X_i^{ac}X_j^{ca}+X_i^{\bar{a}\bar{c}}X_j^{\bar{c}\bar{a}}+X_i^{ca}X_j^{ac}+X_i^{\bar{c}\bar{a}}X_j^{\bar{a}\bar{c}},\nonumber\\
2 (\frac{1}{4}n_i n_j+S^z_iS^z_j) (T^+_iT^-_j + T^-_iT^+_j) &=&
X_i^{ac}X_j^{\bar{c}\bar{a}}+X_i^{\bar{a}\bar{c}}X_j^{ca}+X_i^{ca}X_j^{\bar{a}\bar{c}}+X_i^{\bar{c}\bar{a}}X_j^{ac},\nonumber\\
  (S^+_iS^-_j + S^-_iS^+_j) (T^+_iT^-_j + T^-_iT^+_j) &=&
X_i^{a\bar{c}}X_j^{\bar{c}a}+X_i^{\bar{a}c}X_j^{c\bar{a}}+X_i^{c\bar{a}}X_j^{\bar{a}c}+X_i^{\bar{c}a}X_j^{a\bar{c}},\nonumber\\
4 (\frac{1}{4}n_i n_j-S^z_iS^z_j) (\frac{1}{4}n_i n_j+T^z_iT^z_j) &=& X_i^{aa}X_j^{\bar{a}\bar{a}}+X_i^{\bar{a}\bar{a}}X_j^{aa}+X_i^{cc}X_j^{\bar{c}\bar{c}}+X_i^{\bar{c}\bar{c}}X_j^{cc},\nonumber\\
2 (\frac{1}{4}n_i n_j-S^z_iS^z_j) (\frac{1}{2}n_i+T^z_i) (\frac{1}{2}n_j+T^z_j)&=&
X_i^{aa}X_j^{\bar{a}\bar{a}}+X_i^{\bar{a}\bar{a}}X_j^{aa},\nonumber\\
2 (\frac{1}{4}n_i n_j-S^z_iS^z_j) (\frac{1}{2}n_i+T^z_i) (\frac{1}{2}n_j+T^z_j)&=&
X_i^{aa}X_j^{\bar{a}\bar{a}}+X_i^{\bar{a}\bar{a}}X_j^{aa},\nonumber\\
  (S^+_iS^-_j + S^-_iS^+_j) (\frac{1}{2}n_i+T^z_i) (\frac{1}{2}n_j+T^z_j)&=&
X_i^{a\bar{a}}X_j^{\bar{a}a}+X_i^{\bar{a}a}X_j^{a\bar{a}},\nonumber\\
2 (\frac{1}{4}n_i n_j-S^z_iS^z_j) (\frac{1}{2}n_i-T^z_i) (\frac{1}{2}n_j-T^z_j)&=&
X_i^{cc}X_j^{\bar{c}\bar{c}}+X_i^{\bar{c}\bar{c}}X_j^{cc},\nonumber\\
  (S^+_iS^-_j + S^-_iS^+_j) (\frac{1}{2}n_i-T^z_i) (\frac{1}{2}n_j-T^z_j)&=&
X_i^{c\bar{c}}X_j^{\bar{c}c}+X_i^{\bar{c}c}X_j^{c\bar{c}}.
\end{eqnarray}

Then, the low-energy Hamiltonian
\begin{eqnarray}
 H&=&H^{(ab)} + H^{(bc)} + H^{(ca)} \nonumber\\
 H^{(ca)}&=&H^{(ca)}_{U'-J}+H^{(ca)}_{U'+J}+H^{(ca)}_U \nonumber\\
 H^{(ca)}_{U'-J} = &-& \frac{2}{U'-J}\sum\limits_{ij}
 {\left(\frac{3}{4}n_i n_j + \mathbf{S}_i\cdot\mathbf{S}_j\right)} \nonumber\\
 &\times& \left[ \left({t^{aa}_{ij}}^2 + {t^{cc}_{ij}}^2 \right) \left(\frac{1}{4}n_i n_j - T^z_i T^z_j\right)
 - t^{aa}_{ij} t^{cc}_{ij} \left(T^+_i T^-_j + T^+_i T^-_j\right)\right], \\
 H^{(ca)}_{U'+J} = &-& \frac{2}{U'+J}\sum\limits_{ij}
 {\left(\frac{1}{4}n_i n_j - \mathbf{S}_i\cdot\mathbf{S}_j\right)} \nonumber\\
 &\times& \left[ \left({t^{aa}_{ij}}^2 + {t^{cc}_{ij}}^2 \right) \left(\frac{1}{4}n_i n_j - T^z_i T^z_j\right)
 + t^{aa}_{ij} t^{cc}_{ij} \left(T^+_i T^-_j + T^+_i T^-_j\right)\right], \\
 H^{(ca)}_{U} = &-& \frac{4}{U}\sum\limits_{ij}
 {\left(\frac{1}{4}n_i n_j - \mathbf{S}_i\cdot\mathbf{S}_j\right)} \nonumber\\
 &\times& \left[ {t^{aa}_{ij}}^2 \left(\frac{1}{2}n_i + T^z_i\right) \left(\frac{1}{2}n_j + T^z_j\right)
 + {t^{cc}_{ij}}^2 \left(\frac{1}{2}n_i - T^z_i\right) \left(\frac{1}{2}n_j - T^z_j\right)\right],
  \label{Heff}
\end{eqnarray}
where $H_{U'-J}^{(ca)}$ results from the virtual processes with $S=1$ intermediate states. On the other hand, $H_{U'+J}^{(ca)}$ and $H_U^{(ca)}$ result from the virtual processes with interorbital and intraorbital $S=0$ intermediate states, respectively.

For $t^{aa}_{ij}=t^{cc}_{ij}=t_{ij}$,
\begin{eqnarray}
 H_{U'-J}^{(ca)} = &-& \frac{4}{U'-J}\sum\limits_{ij}{{t_{ij}}^2
     \left(\frac{3}{4}n_i n_j + \mathbf{S}_i\cdot\mathbf{S}_j\right)
     \left(\frac{1}{4}n_i n_j - \mathbf{T}_i\cdot\mathbf{T}_j\right)}, \\
 H_{U'+J}^{(ca)} = &-& \frac{4}{U'+J}\sum\limits_{ij}{{t_{ij}}^2
     \left(\frac{1}{4}n_i n_j - \mathbf{S}_i\cdot\mathbf{S}_j\right)
     \left(\frac{1}{4}n_i n_j + \mathbf{T}_i\cdot\mathbf{T}_j - 2T^z_i T^z_j\right)}, \\
 H_{U}^{(ca)} = &-& \frac{4}{U}\sum\limits_{ij}
 {{t_{ij}}^2 \left(\frac{1}{4}n_i n_j - \mathbf{S}_i\cdot\mathbf{S}_j\right)
 \left(\frac{1}{2}n_i n_j + 2 T^z_i T^z_j\right)}.
  \label{Hsimplet}
\end{eqnarray}
For $t^{aa}_{ij}=t^{cc}_{ij}=t_{ij}$, $J=0$, $U=U'$, one obtains the symmetric result:
\begin{eqnarray}
 H^{(ca)} = - \frac{4}{U}\sum\limits_{ij}
{t_{ij}}^2 &\Bigg[& \left(\frac{3}{4}n_i n_j + \mathbf{S}_i\cdot\mathbf{S}_j\right)
     \left(\frac{1}{4}n_i n_j - \mathbf{T}_i\cdot\mathbf{T}_j\right) \nonumber\\
&+& \left(\frac{1}{4}n_i n_j - \mathbf{S}_i\cdot\mathbf{S}_j\right)
     \left(\frac{3}{4}n_i n_j + \mathbf{T}_i\cdot\mathbf{T}_j\right) \Bigg].
  \label{Hsimple}
\end{eqnarray}

\end{widetext}


\begin{thebibliography}{99}
\bibitem{khomskii} S. V. Streltsov and D. I. Khomskii, Phys. Rev. B \textbf{77}, 064405 (2008).
\bibitem{brink} M. J. Konstantinovi\'{c}, J. van den Brink, Z. V. Popovi\'{c}, V. V. Moshchalkov, M. Isobe and Y. Ueda, Phys. Rev. B {\bf 69},  020409 (2004).
\bibitem{brinkEPL} J. van Wezel and J. van den Brink, Europhys. Lett. {\bf 75}, 957 (2006).
\bibitem{motome} T. Hikihara and Y. Motome, J. Phys. Soc. Jap. Suppl. {\bf 74}, 212 (2005).
\bibitem{JPSJ} M. Isobe, E. Ninomiya, A. N. Vasili'ev and Y. Ueda, J. Phys. Soc. Jap. {\bf 71}, 1423 (2002).
\bibitem{spingap} H. J. Silverstein et. al., Phys. Rev. B {\bf 90}, 140402 (2014).
\bibitem{popovic} Z. S. Popovi\'{c}, \v{Z}eljko V. \v{S}ljivan\v{c}anin, and Filip R. Vukajlovi\v{c}, Phys. Rev. Lett. \textbf{93}, 036401 (2004).
\bibitem{khomskiiPRL} S. V. Streltsov, O. A. Popova, and D. I. Khomskii, Phys. Rev. Lett. \textbf{96}, 249701 (2006).
\bibitem{mila} F. Mila and F.-C. Zhang, Eur. Phys. J. B \textbf{16} 7 (2000).
\bibitem{jackeli} G. Jackeli and G. Khaliullin, Phys. Rev. Lett. \textbf{102}, 017205 (2009).
\bibitem{SI} See online Supplemental Material.
\bibitem{yin_prb09} W.-G. Yin and W. Ku, Phys. Rev. B \textbf{79}, 214512 (2009).
\bibitem{ep} It can be gapped by electron-phonon interaction that utilizes the $2k_F$ instability to drive the Peierls transition, in which the unit cell is doubled to have six Ti atoms, compatible with Fig.~\ref{fig:relabel}(d).
\bibitem{Manmana2011} Salvatore R. Manmana, Kaden Hazzard, Gang Chen, Adrian E. Feiguin and Ana Maria Rey, Phys. Rev. A {\bf 84}, 043601 (2011).
\bibitem{Sutherland} B. Sutherland, Phys. Rev. A{\bf 5}, 1372 (1972).
\bibitem{White2004} S. R. White and A. E. Feiguin, Phys. Rev. Lett. {\bf 93} 076401 (2004).
\bibitem{Daley2004} A. Daley, C. Kollath, U. Schollw\"ock and G. Vidal, J. Stat. Mech: Theor. Phys. P04005 (2004).
\bibitem{vietri} A.~E. Feiguin, in {\it XV Training Course in the Physics of Strongly Correlated Systems}, Vol. 1419 (AIP Proceedings, 2011), p. 5.
\bibitem{Paeckel2019} S. Paeckel, T. Kahler, A. Swoboda, S. R. Manmana, U. Schollw\"ock and C. Hubig, arXiv:1901.05824.
\bibitem{White2008} Steven R. White and Ian Affleck, Phys. Rev. B {\bf 77}, 134437 (2008).
\end{thebibliography}

\begin{references}
\bibitem{wien2k} K. Schwarz, P. Blaha, and G. K. H. Madsen, Comput. Phys. Commun. \textbf{147}, 71 (2002).
\bibitem{gga} J. P. Perdew, K. Burke, and M. Ernzerhof, Phys. Rev. Lett. \textbf{77}, 3865 (1996).
\bibitem{ku_prl02} W. Ku et al., Phys. Rev. Lett. 89, 167204 (2002).
\bibitem{yin_prl06} W.-G. Yin, D. Volja, and W. Ku, Phys. Rev. Lett. \textbf{96}, 116405 (2006).
\bibitem{yin_prl06_TMD} R. L. Barnett, A. Polkovnikov, E. Demler, W.-G. Yin, and W. Ku, Phys. Rev. Lett. \textbf{96}, 026406 (2006).
\bibitem{yin_prb09} Wei-Guo Yin and Wei Ku, Phys. Rev. B \textbf{79}, 214512 (2009).
\end{references}

\end{document}